# Multi-channel end-to-end neural network for speech enhancement, source localization, and voice activity detection

Yuan Chen[1]; Yicheng Hsu[2]; Mingsian R. Bai[3]

[1] Department of Power Mechanical Engineering, National Tsing Hua University, Taiwan

[2] Department of Power Mechanical Engineering, National Tsing Hua University, Taiwan

[3] Electrical Engineering, National Tsing Hua University, Taiwan

**ABSTRACT**

Speech enhancement and source localization has been active research for several decades with a wide range of real-world applications. Recently, the Deep Complex Convolution Recurrent network (DCCRN) has yielded impressive enhancement performance for single-channel systems. In this study, a neural beamformer consisting of a beamformer and a novel multi-channel DCCRN is proposed for speech enhancement and source localization. Complex-valued filters estimated by the multi-channel DCCRN serve as the weights of beamformer. In addition, a one-stage learning-based procedure is employed for speech enhancement and source localization. The proposed network composed of the multi-channel DCCRN and the auxiliary network models the sound field, while minimizing the distortionless response loss function. Simulation results show that the proposed neural beamformer is effective in enhancing speech signals, with speech quality well preserved. The proposed neural beamformer also provides source localization and voice activity detection (VAD) functions.

Keywords: multi-channel speech enhancement, source localization, deep learning

## 1. INTRODUCTION

The goal of speech enhancement is to extract the target speech from the noisy signal. Since interfering noise and reverberation are pervasive in real-world, speech enhancement is essential in many applications to generate the enhanced speech waveform. Recently, deep neural network (DNN) has achieved impressive results in speech enhancement problems. Monaural processing with DNN is well-known approach to speech enhancement, which exploits the information in time domain [1-3] or time-frequency domain [4-6]. Time-frequency masking-based approaches in [4,5] employing supervised learning are proved to be effective in speech enhancement problems. Based on the success in time-frequency masking approach, DCCRN that exploits complex operations performs competitively over other previous networks [6].

However, in far-field applications, such as hands-free teleconferencing and smart speaker, the enhancement performance often degrades due to interference, reverberation, noise, etc. Motivated by fabulous results in DNN-based monaural speech enhancement, several multi-channel DNN-based speech enhancement approaches have been studied. Multi-channel methods benefit from spatial features given by inter-channel information. Employing multi-channel signals, a typical strategy is to combine DNN with conventional beamforming methods. In [7], DNN estimates time-frequency (T-F) masks to specify the dominant T-F bins for the signal of interest and noise. T-F masks are used to calculate the spatial covariance matrix for beamformer's weights such as minimum variance distortionless response (MVDR) [8] and generalized eigenvalue (GEV) [9]. [10] proposes an all deep learning MVDR (ADL-MVDR) beamformer, where the inverse operation in the aforementioned DNN method is unnecessary. Without reference to conventional beamformers, filter-based approachs applying filter-and-sum beamformer are proposed either in time domain [11] and time-frequency domain [12-14]. Previous studies have shown the importance of using time-frequency domain beamforming techniques in sensor array systems [15]. A state-of-the-art multiple-in-multiple-out (MIMO) U-net neural beamformer in time-frequency domain is proposed in [14] to produce a complex beamformer. The U-net neural beamformer utilizes spatial features implicitly by analyzing spectral features. This system ranked first in the ConferencingSpeech 2021 Challenge.

---

[1] cya10230223@gmail.com
[2] shane.ychsu@gapp.nthu.edu.tw
[3] msbai@pme.nthu.edu.tw

In this study, inspired by the concept in the U-net beamformer [14] and DCCRN [6], the MIMO-DCCRN neural beamformer is proposed. In addition, a one-stage learning based procedure based on MIMO-DCCRN is formulated for both speech enhancement and localization. The neural beamformer incorporates the traditional beamforming structure with a DNN model by adding a filter-and-sum operation at the end of the beamformer, similar to the U-net structure in [14]. The complex encoder in MIMO-DCCRN extracts the feature from both spectral and spatial perspectives. The framework is then equipped with complex long short-term memory (LSTM) in the bottleneck to model temporal context. The end of the network is complex decoder to estimate complex spatial filters. Comparing to the existing neural beamformers, the proposed neural beamformer deliberates the complex operation additionally. Further, depending on complex filters estimation, two localization solutions are derived from distortionless constraint. The first solution is to maximize the magnitude of steered beamformer response with free-field steering vectors [8] across the selected zones. Second, combine MIMO-DCCRN with the auxiliary network which models the sound field in the proposed network. Accordingly, the loss function is formulated to correlate with the enhancement quality and localization accuracy given distortionless constraint.

Note that in contrast to conventional beamformers such as MVDR [8], that the second-order spatial statistics and accurate direction of arrival (DOA) require to be explicitly calculated, our approaches directly estimate beamformer's weights and further locate the position of the target speaker. VAD can also be conducted from localization results. Consequently, the proposed network is able to handle enhancement, localization and VAD simultaneously. The simulation results show that this novel neural beamformer outperforms the existing DCCRN and MIMO U-net in enhancement. It is even shown that multi-tasking training, including speech enhancement and localization, improves enhancement in perceptual quality and intelligibility.

## 2. PROBLEM FORMULATION

Given a $M$-microphone time-frequency domain signal $\mathbf{y}(t,f) = [Y_0(t,f) \ldots Y_{M-1}(t,f)]^T \in \mathbb{C}^{M \times 1}$ recorded in a reverberant and noisy environment, where $t = 1,\ldots,T$ denotes the time frame, $f = 1,\ldots,F$ denotes the frequency bin, the signal model in the short-time Fourier transform (STFT) domain is formulated as:

$$\mathbf{y}(t,f) = \mathbf{s}(t,f) + \mathbf{h}(t,f) + \mathbf{n}(t,f), \tag{1}$$

where $\mathbf{s}(t,f) = [S_1(t,f) \ldots S_M(t,f)]^T \in \mathbb{C}^{M \times 1}$ denotes the direct sound and early reflection components of a single target speech received by microphones, and $\mathbf{h}(t,f)$, $\mathbf{n}(t,f) \in \mathbb{C}^{M \times 1}$ denotes late reverberation components and reverberant noises, respectively. (1) can be expressed as

$$\mathbf{y}(t,f) = \mathbf{s}(t,f) + \mathbf{v}(t,f), \tag{2}$$

where $\mathbf{v}(t,f) = \mathbf{h}(t,f) + \mathbf{n}(t,f)$ is the undesired signal. The proposed approach estimates the signal of interest $\hat{S}(t,f)$ by filter-and-sum beamforming:

$$\hat{S}(t,f) = \sum_{m=0}^{M-1} \{w_m^*(t,f) \cdot Y_m(t,f)\} = \mathbf{w}^H(t,f)\mathbf{y}(t,f). \tag{3}$$

Thus, the neural beamformer system aims to estimate complex filter weights $\mathbf{w}(t,f) = [w_0(t,f) \ldots w_{M-1}(t,f)] \in \mathbb{C}^{M \times 1}$ for $M$ microphones.

## 3. THE PROPOSED SYSTEM

Our proposed framework is illustrated in Fig. 1(a). The system is composed of an enhancement and a localization module. The signal from each channel is transformed into time-frequency domain representation as the input of our system. These features first passed through MIMO-DCCRN (Sec

3.1), which is the enhancement module, to generate filter coefficients. The estimated filter weights are then fed to the localization module (Sec 3.2) to analyze information from each channels and output localization results.

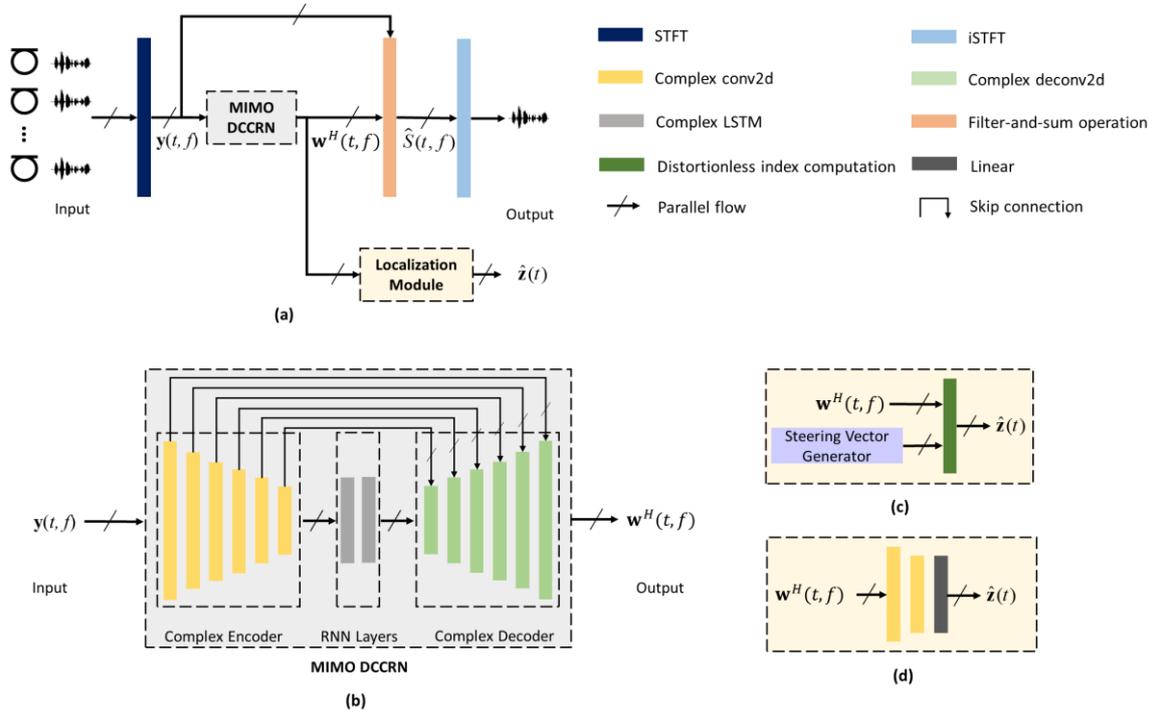

Figure 1 – Block diagram of (a) the proposed system, (b) MIMO-DCCRN, (c) the signal processing localization module (SPLM), and (d) the neural localization module (NLM).

### 3.1 MIMO DCCRN

DCCRN in [6] is the complex network modified by CRN in [5]. In this study, we extend DCCRN to MIMO-DCCRN. Signals captured by different channels in time-frequency domain through STFT are taken as the MIMO-DCCRN input. The real and imaginary parts of complex are stacked together to obtain a 2-channel tensor of size $2 \times T \times F$, where $T$ represents the time steps and $F$ represents the number of frequency bins, which are used as input features. Considering all $M$ channels, it leads to a $2M \times T \times F$ dimensional input $\mathbf{y}$ to the framework. The complex encoder consists of complex convolution layers followed by a complex batch normalization (BN) and PReLU activation function. In the multi-input complex encoder, the input across different channels is interpreted both in spectral and spatial aspects. Features extracted by the complex encoder are utilized in complex LSTM to be further interpreted in temporal information. The complex decoder is designed to be the structure symmetrical to the complex encoder to estimate $\mathbf{w}^H$ based on the above feature extraction. Skip-connection is conductive to flowing the gradient by concentrating the complex encoder and decoder. Fig. 1(b) illustrates the proposed architecture for multi-channel speech enhancement.

### 3.2 Localization and VAD

Aiming to locate the target speaker position, complex filters estimated in the aforementioned system are passed through the localization module. The speech enhancement and localization problems are jointly tackled via multi-task learning. As shown in Fig. 1(a), a localization module is cascaded with the output of the MIMO-DCCRN. The main goal of an array beamformer is to restore the source signal from a target direction as a spatial filter, while suppressing noise and interference from other directions. To achieve this goal, the distortionless constraint is imposed to maintain a constant response for the target source direction. This motivates the proposed localization module to utilize the following distortionless constraint as a localization criterion:

$$\mathbf{w}^H \mathbf{a} = 1, \tag{4}$$

where $\mathbf{a}$ is a steering vector pertaining to the target speaker direction.

### 3.2.1 The Signal Processing Localization Module (SPLM)

The signal processing localization module (SPLM) is illustrated in Fig. 1 (c). By assuming $N$ candidate zones in azimuthal angles, the free-field plane wave steering vector $\mathbf{a}_n(f) \in \mathbb{C}^{M \times 1}$, $n = 1, \ldots, N$ is predesignated according to the pre-specified angular grid, which is the center of zone-$n$,

$$\mathbf{a}_n(f) = \begin{bmatrix} e^{-j\mathbf{k}_n \cdot \mathbf{r}_1} & e^{-j\mathbf{k}_n \cdot \mathbf{r}_2} & \cdots & e^{-j\mathbf{k}_n \cdot \mathbf{r}_M} \end{bmatrix}^T, \tag{5}$$

where $\mathbf{r}_m$ is the position vector of the $m$th microphone, $\mathbf{k}_n = -k\boldsymbol{\kappa}_n = -(\omega/c)\boldsymbol{\kappa}_n = -(2\pi f/c)\boldsymbol{\kappa}_n$ is the wave vector, $\boldsymbol{\kappa}_n$ denotes the direction of arrival (DOA) vector that is a unit vector pointing to the look direction, $k$ denotes the wave number, and $c$ is the speed of sound. Here we define the estimated localization mapping be $\hat{\mathbf{z}}(t) = [\hat{z}_1(t) \ \ldots \ \hat{z}_N(t)] \in \mathbb{R}^N$, which can be viewed as probabilities, composed of distortionless index. The distortionless index $\hat{z}_n(t)$ of the $n$-direction and time frame $t$ is formulated based on the estimated filter coefficients $\mathbf{w}^H$ in MIMO-DCCRN and predefined steering vectors,

$$\hat{z}_n(t) = \frac{1}{F} \sum_{f=1}^{F} \left| \mathbf{w}^H(t,f) \mathbf{a}_n(f) \right|. \tag{6}$$

The active zone can be determined by peak finding of $\hat{z}_n(t)$

$$\hat{n}(t) = \arg\max_n \hat{z}_n(t). \tag{7}$$

In addition, the peak of $\hat{z}_n(t)$ give the information of voice activity

$$\text{VAD}(t) = \max_n \hat{z}_n(t). \tag{8}$$

Hereafter, we denote MIMO-DCCRN-SPLM-$N$ as the signal processing localization module with predefined $N$ zones cascaded with MIMO-DCCRN.

### 3.2.2 The Neural Localization Module (NLM)

In Eq. (5), steering vectors are derived based on the free-field plane wave assumption, which does not well accommodate in the real-world application. Thus, the auxiliary network is further proposed to model the realistic sound field. The neural network consisting of the complex convolutional layer with complex BN, PReLU activation and the linear layer with sigmoid activation function replaces coefficients of steering vector. As shown in Fig. 1(d), filter coefficients are processed through the localization network to estimate $\hat{\mathbf{z}}(t)$. The direction and VAD are computed according to Eq. (7) and Eq. (8), respectively. Hereafter, the localization module employing the learning-based sound field with predefined $N$ zones for localization in training and cascaded with the MIMO-DCCRN is denoted as MIMO-DCCRN-NLM-$N$.

### 3.3 Loss Functions

In this study, different loss functions are introduced for enhancement and localization problems. For speech enhancement, the objective is to minimize the negative scale-invariant source-to-noise ratio (SI-SNR), which has commonly used as an evaluation metric for enhancement replacing the mean square error (MSE) and signal-to-distortion ratio (SDR) [16]. SI-SNR is defined as

$$\begin{aligned} \mathbf{s}_{\text{target}} &= \frac{\langle \hat{\mathbf{s}}, \mathbf{s} \rangle \mathbf{s}}{\|\mathbf{s}\|_2^2} \\ \text{SI-SNR} &= 20\log_{10} \frac{\|\mathbf{s}_{\text{target}}\|_2^2}{\|\hat{\mathbf{s}} - \mathbf{s}_{\text{target}}\|_2^2} \end{aligned}, \tag{9}$$

where $\hat{\mathbf{s}}$ and $\mathbf{s}$ are the estimated and original clean time-domain waveform, respectively. $\langle \cdot, \cdot \rangle$ denotes the dot product between two vectors and $\|\cdot\|_2$ is Euclidean norm. $loss_{\text{SI-SNR}}$ is denoted as the average of negative SI-SNR in time index.

For DOA estimation, if the target source is activated at azimuth angle $\theta$ and time frame $t$, the ground truth of localization mapping is defined as $\mathbf{z}(t) = [z_1(t) \ \ldots \ z_N(t)] \in \mathbb{R}^N$, where

$$z_n(t) = \begin{cases} 1, & -\frac{360°}{2N} + (n-1) \cdot \frac{360°}{N} < \theta \leq -\frac{360°}{2N} + n \cdot \frac{360°}{N} \\ 0, & \text{otherwise} \end{cases} \quad (10)$$

While the target source is not activated, $z_n(t) = 0$, $n=1,\ldots,N$. The localization problem is formulated as a classification task with $\mathbf{z}(t)$ and $\hat{\mathbf{z}}(t)$. As mentioned above, with $z_n(t) \in \{0,1\}$ and $\hat{z}_n(t) \in [0,1]$, the binary cross-entropy is minimized in training:

$$loss_{\text{BCE}} = -\frac{1}{TN} \sum_{t=1}^{T} \sum_{n=1}^{N} z_n(t) \log \hat{z}_n(t) + (1 - z_n(t)) \log(1 - \hat{z}_n(t)), \quad (11)$$

Therefore, the total loss of the multi-task network is

$$loss = loss_{\text{BCE}} + \gamma loss_{\text{SI-SNR}}, \quad (12)$$

where $\gamma$ is the weighting constant and is set to 1 in this paper.

## 4. SIMULATION STUDY

### 4.1 Dataset Preparation

To validate the proposed system, we conduct a simulation data based on four open datasets. The clean utterances for training and testing are selected from the train-clean-360 and dev-clean subsets of the LibriSpeech corpus [17], which contains utterances from 921 and 40 speakers, respectively. The noisy training and testing data are synthesized with noise corpus using the MS-SNSD dataset [18], FSDnoisy18k dataset [19] and the Free Music Archive (FMA) [20]. From MS-SNSD dataset, noise data excluded babble noise is chosen for directional interference noises. In FSDnoisy18k dataset, noises which are lack of directionality, such as the environment noise, are not considered for data preprocessing in this work. In this paper, we process the audio both contained stationary noises, such as music and engine, and transient noises, such as glass breaking and typing noise.

The duration of mixed noisy audio signals prepared for training and testing is 6-s. Each mixture is composed of 4-s clean speech audio clips, which were randomly inserted into 6-s duration, Gaussian white noise, which is used as sensor noises, and interference noises. A 6-element uniform circular array (UCA) with a radius of 5 cm is employed. In the training phase, the signal-to-noise ratio (SNR) of sensor noise ranged from 10 to 30 dB. The signal-to-interference ratio (SIR) of inference noise is selected from -5 to 15 dB. The room size, the distance between sound source and microphone array, and T60 are sampled from $\{4 \times 4 \times 3, 5 \times 5 \times 3, 6 \times 6 \times 3\}$ m$^3$, $\{[1, 1.5], [1, 2], [1, 2.5]\}$ m, and $\{[0.16, 0.32], [0.32, 0.48], [0.48, 0.64]\}$ second, respectively. In the test dataset, the room size and T60 are selected to be $5 \times 5 \times 3$ m$^3$ and 0.32-s. The distance between the target speaker and the microphone array is fixed to 1 m, and the distance between the interference noise and the microphone array is fixed to 2 m. In the simulation, the microphone array is placed at the center of the room. The target speaker is placed sequentially on a semicircle centered at the microphone, from 0° to 180° with 1° increments, whereas the interference is located sequentially on a semicircle centered at the microphone, from 180° to 360° at increments of 1°. The noisy signals received by the microphone array are generated by convolving the anechoic clean signals with the respective room impulse responses (RIRs) simulated using the image-source method [21]. We generate 37000, 460 and 600 mixtures for the training, validation, and test dataset, respectively.

### 4.2 Training Setup and Baselines

All the waveforms are sampled at 16 kHz. The STFT settings are a Hanning window with 25-ms length, a 6.25-ms hop size, and a 512-point fast Fourier transform. In the training stage, Adam optimizer is adopted with the learning rate as 0.001. All models are trained for 100 epochs.

In our proposed system, the number of channels of the complex convolution layers in the encoder of MIMO-DCCRN are 16, 32, 64, 128, 256, 256, respectively. The number of channels of the decoder are the same of the encoder in reversed order. All complex (de-)convolution of the encoder (decoder) use $5\times2$ kernel size, with stride $2\times1$. For RNN layers, 256 hidden units are applied in complex LSTM. In NLM, the complex convolution layers are $2N$, $2N$, and the joined linear layers are 32, 1. In the following, we compare the improvements of MIMO-DCCRN to simply enhance speech, MIMO-DCCRN-SPLM-12 and MIMO-DCCRN-NLM-12 which are combining localization on 12 zones, and MIMO-DCCRN-SPLM-36 and MIMO-DCCRN-NLM-36 to further localize on 36 zones for higher accuracy.

Two baselines are adopted in the following, including MIMO U-net [14] and DCCRN [6]. MIMO U-net is the multi-channel speech enhancement system that achieved impressive performance in neural beamforming. DCCRN outperforms other speech enhancement methods in the monaural scenario.

### 4.3 Results

#### 4.3.1 Speech Enhancement Performance

To quantify the signal enhancement performance, perceptual evaluation of speech quality (PESQ) [22] and short-time objective intelligibility (STOI) [23] are employed. Table 1 summarizes the PESQ and STOI scores for the test dataset. First, we examine the effects of the novel MIMO-DCCRN on the level of performance improvement. Table 1 shows that our proposed methods yield better performance than baseline methods. Comparing MIMO-DCCRN and DCCRN, multi-channel speech enhancement attains significantly improvements than monaural enhancement, especially the STOI score at low SIR. For multi-channel enhancement, the MIMO-DCCRN system consisting of complex operations achieves great improvement than MIMO U-net. The results proved that MIMO-DCCRN is more robust to scenarios at different SIR.

Second, we examine the effects of combining localization to speech enhancement. According to Table 1, MIMO-DCCRN-NLM-36 performs the best among all methods. On the contrary, PESQ and STOI decrease slightly in the MIMO-DCCRN-SPLM system without learning-based sound field. NLM thus is proved to be more reliable than SPLM. Furthermore, MIMO-DCCRN-SPLM-12 and MIMO-DCCRN-NLM-12, which using small numbers of zones in localization, degrade the enhanced signal quality, since insufficient number of zones designed in localization leads to misalignment in neural beamforming. The result suggests that appropriate localization module aids beamforming in maintaining the desired signal at specific direction. Thus, MIMO-DCCRN-NLM-36 can further improve the enhanced signal quality.

Table 1 – Speech enhancement performance

| Performance Metric | PESQ | | | | | STOI (in %) | | | | |
|---|---|---|---|---|---|---|---|---|---|---|
| SIR (dB) | -10 | -5 | 0 | 10 | Avg. | -10 | -5 | 0 | 10 | Avg. |
| Noisy | 1.58 | 1.68 | 1.85 | 2.03 | 1.79 | 70.81 | 75.85 | 82.26 | 86.29 | 78.80 |
| DCCRN | 1.69 | 1.86 | 2.12 | 2.38 | 2.01 | 72.75 | 78.94 | 85.94 | 89.42 | 81.76 |
| MIMO U-net | 1.90 | 2.08 | 2.38 | 2.63 | 2.25 | 77.93 | 82.48 | 87.91 | 90.66 | 84.75 |
| MIMO-DCCRN | **2.43** | **2.71** | **3.07** | **3.31** | **2.88** | 86.64 | 90.63 | 94.56 | 96.07 | 91.98 |
| MIMO-DCCRN-SPLM-12 | 2.33 | 2.61 | 2.98 | 3.25 | 2.79 | 86.26 | 90.37 | 94.43 | 95.96 | 91.76 |
| MIMO-DCCRN-NLM-12 | 2.30 | 2.59 | 2.95 | 3.21 | 2.76 | 86.40 | 90.58 | 94.58 | 96.05 | 91.90 |
| MIMO-DCCRN-SPLM-36 | 2.33 | 2.59 | 2.93 | 3.19 | 2.76 | 86.31 | 90.37 | 94.32 | 95.91 | 91.73 |
| MIMO-DCCRN-NLM-36 | **2.43** | **2.71** | **3.07** | **3.31** | **2.88** | **86.94** | **90.87** | **94.66** | **96.11** | **92.15** |

#### 4.3.2 Localization and VAD Performance

To evaluate the performance of the proposed DOA estimation, we define three evaluation metrics: total accuracy (ACC), adjacent error rate (AER), and other error rate (OER) in Eq. (12).

$$ACC = \frac{L_{\text{true}}}{L_{\text{all}}}, \; AER = \frac{L_{\text{adjacent}}}{L_{\text{all}}}, \; OER = \frac{L_{\text{false}}}{L_{\text{all}}}, \tag{12}$$

where $L_{\text{true}}$ is the number of true estimations, $L_{\text{adjacent}}$ is false estimations on adjacent zones, $L_{\text{false}}$ is false estimations except on adjacent zones, and $L_{\text{all}}$ is the total number of localization estimations. In this evaluation, we only estimate localization when the target source was activated. The localization results are illustrated in Table 2, where different localization module and different number of zones are compared. As the number of zones increase, the performance of SPLM decrease considerably. This result reveals that NLM is more robust in application than SPLM.

Moreover, Fig. 2 shows how the VAD output has learned from localization module. The system enables the model to learn only the interfering voice activity while ignoring the interfering voice activity. Further evaluation on the quality of the VAD output is beyond the scope of this work.

Table 2 – Localization performance

| Performance Metric | ACC (%) | | | | | AER (%) | | | | | OER (%) | | | | |
|---|---|---|---|---|---|---|---|---|---|---|---|---|---|---|---|
| SIR (dB) | -10 | -5 | 0 | 10 | Avg. | -10 | -5 | 0 | 10 | Avg. | -10 | -5 | 0 | 10 | Avg. |
| MIMO-DCCRN-SPLM-12 | 80.1 | 83.3 | 85.7 | 86.2 | 83.8 | 15.8 | 14.1 | 12.5 | 12.5 | 13.7 | 4.0 | 2.5 | 1.7 | 1.3 | 2.4 |
| MIMO-DCCRN-NLM-12 | **90.9** | **94.1** | **95.6** | **96.1** | **94.2** | **6.8** | **4.7** | **3.8** | **3.4** | **4.7** | **2.2** | **1.1** | **0.7** | **0.5** | **1.1** |
| MIMO-DCCRN-SPLM-36 | 55.0 | 57.4 | 59.4 | 60.4 | 58.1 | 36.8 | 37.1 | 37.3 | 37.1 | 37.1 | 7.7 | 5.1 | 3.1 | 2.4 | 4.6 |
| MIMO-DCCRN-NLM-36 | **80.8** | **85.1** | **88.8** | **90.0** | **86.2** | **15.5** | **12.6** | **9.9** | **9.0** | **11.8** | **3.8** | **2.3** | **1.4** | **1.0** | **2.1** |

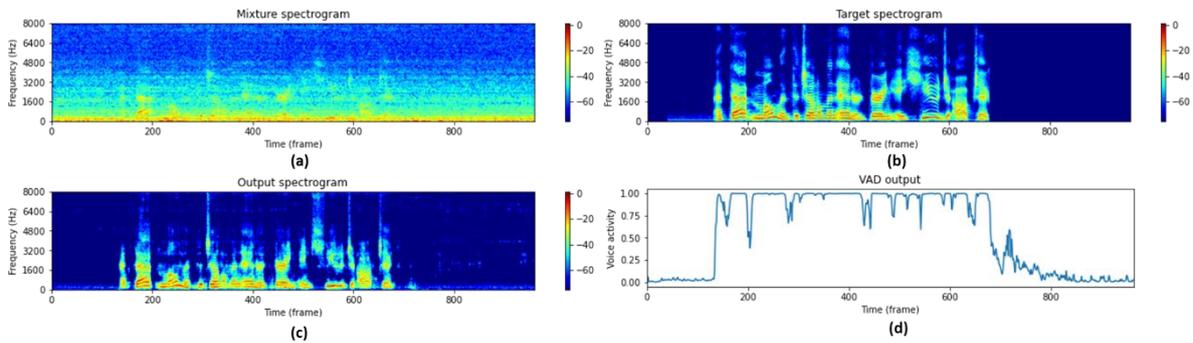

Figure 2 – Spectrogram of (a) noisy mixture signal, (b) clean speech (c) enhanced signal by MIMO-DCCRN-NLM-12, (d) VAD output.

## 5. CONCLUSIONS

In this paper, a DCCRN-based neural beamformer, is proposed for multi-channel speech enhancement and localization. MIMO-DCCRN exhibits superior performance in speech enhancement. By employing distortionless constraint in loss function, the neural beamformer also provides effective localization and VAD functions. Combining neural beamformer and the learning-based sound field for localization with a sufficiently fine grid proves useful in enhancing speech quality. In brief, the combined MIMO-DCCRN-NLM-36 system has achieved significantly improved performance in speech enhancement, as compared to baselines.

## ACKNOWLEDGEMENTS

This work was supported by the Ministry of Science and Technology (MOST), Taiwan, under the project number 110-2221-E-007-027-MY3.